\documentclass[useAMS,usenatbib]{mn2e}

\usepackage{amssymb}
\usepackage{graphicx}
\usepackage{color}
\usepackage{amsmath}

\title[MWDs with debris discs]{Magnetic white dwarfs with debris discs}
\author[B. K\"ulebi et al.]
       {B. K\"ulebi$^{1,2}$\thanks{E-mail: kulebi@ice.cat}, 
        K.Y. Ek{\c s}i$^{3}$\thanks{E-mail: eksi@itu.edu.tr}, 
        P. Lor{\'e}n--Aguilar$^{4}$, 
        J. Isern$^{1,2}$ and
        E. Garc{\'\i}a--Berro$^{2,5}$\\
	   $^{1}$ Institut de Ci\`encies de l'Espai (CSIC), 
           Facultat de Ci\`encies, 
           Campus UAB, 
           Torre C5-parell, 
           08193 Bellaterra, 
           Spain\\
	   $^{2}$ Institute for Space Studies of Catalonia,
           c/Gran Capit\`a 2--4, Edif. Nexus 104,   
           08034  Barcelona, 
	   Spain\\
	   $^{3}$ Istanbul Technical University, 
	   Faculty of Science and Letters,
	   Physics Engineering Department,
	   Maslak 34469, Istanbul, Turkey\\
	   $^{4}$ University of Exeter, 
	   School of Physics, 
	   Stocker Road, 
           Exeter EX4 4QL,
	   United Kingdom\\
	   $^{5}$ Departament de F\'\i sica Aplicada, 
           Universitat Polit\`ecnica de Catalunya,  
           c/Esteve Terrades, 5,  
           08860 Castelldefels,  
           Spain\\
	   }
\begin{document}

\def\Rej{\mbox{RE{}\,J~0317$-$853}}
\newcommand{\Teff}{\hbox{$T_{\rm eff}$}}
\newcommand{\Msolar}{\mbox{\,${\rm M}_{\sun}$}}
\newcommand{\Rsolar}{\mbox{\,${\rm R}_{\sun}$}}
\newcommand{\Mch}{\mbox{\,$M_{\rm Ch}$}}
\newcommand{\M}{\mbox{M}}
\newcommand{\RWD}{\mbox{$R_{\rm WD}$}}
\newcommand{\MWD}{\mbox{$M_{\rm WD}$}}
\def\euve{\mbox{WD{}\,1658{}+{}441}}
\def\PGa{\mbox{PG{}\,1015{}+{}014}}
\def\PGb{\mbox{PG{}\,1031{}+{}234}}
\def\Grw{\mbox{Grw~$+70^\circ 8247$}}

\def\apj{ApJ}
\def\apjs{ApJS}
\def\apjl{ApJL}
\def\nat{Nature}
\def\aj{AJ}
\def\aap{A\&A}
\def\mnras{MNRAS}
\def\pasj{PASJ}
\def\nar{New Astron. Rev.}
\def\araa{ARAA}
\def\actaa{Acta Astron.}
\def\pasp{PASP}
\def\na{New Astron.}
\def\aapr{AAPR}

\date{\today}

\maketitle

\begin{abstract}
It has long been accepted that a possible mechanism for explaining the
existence of  magnetic white  dwarfs is the  merger of a  binary white
dwarf system, as there are viable mechanisms for producing sustainable
magnetic fields within the merger product.  However, the lack of rapid
rotators  in  the magnetic  white  dwarf  population  has been  always
considered a  problematic issue  of this scenario.   Smoothed Particle
Hydrodynamics simulations show that in  mergers in which the two white
dwarfs have different masses a  disc around the central compact object
is formed.  If  the central object is magnetized  it can interact with
the disc  through its magnetosphere.   The torque applied by  the disc
changes the spin of the star, whereas the transferred angular momentum
from the  star to the disc  determines the properties of  the disc. In
this work we build a model  for the disc evolution under the effect of
magnetic  accretion, and  for the  angular momentum  evolution  of the
star, which can be compared  with the observations. Our model predicts
that  the magnetospheric  interaction  of magnetic  white dwarfs  with
their discs results  in a significant spin down, and  we show that for
magnetic white  dwarfs with  relatively strong fields (larger than $10\,$MG) the
observed rotation  periods of the magnetic white  dwarf population can
be  reproduced.   We  also   investigate  whether  turbulence  can  be
sustained during the late phases of the evolution of the system.  When
a  critical temperature  below which  turbulence is  not  sustained is
introduced  into  the  model,    the periods  of  the  three  fast
rotating,  strongly  magnetic,  massive  white  dwarfs  in  the  solar
neighborhood are recovered.
\end{abstract}

\begin{keywords}
stars: white dwarfs, stars: magnetic field, accretion discs
\end{keywords}

\section{Introduction}
\label{sec:introduction}

Magnetic  white dwarfs (MWDs)  --- white  dwarfs with  field strengths
ranging from about 1~kG  \citep{Jordanetal07} up to approximately 1~GG
\citep{Kawkaetal07, Kulebietal09} --- comprise more than about 10\% of
all  white dwarfs.   There are  two possibilities  to account  for the
observed magnetic  fields.  According to the  fossil field hypothesis,
these white  dwarfs descend from Ap/Bp  stars \citep{Angeletal81}, and
their magnetic fields are remnants  of the previous evolution.  One of
the  key   properties  of the  population  of MWDs  is their  mass
distribution,  since it  turns out  that their  average mass  is 
considerably   larger    than   that   of    the   field   population
\citep{Kawkaetal07}.   Moreover, the  reverse  is also  true, and  the
incidence of  magnetism is larger  in the population of  massive white
dwarfs \citep{VennesKawka08}.   Within the framework of the fossil
field hypothesis  this has been explained  as a result  of the average
higher  mass  of  the  magnetic progenitor  systems  Ap/Bp.   However,
detailed population synthesis studies show that the observed number of
Ap/Bp stars  is insufficient  to explain the  number of MWDs,  and the
contribution of  an unseen population  of weakly magnetized  A/B stars
needs  to be  invoked  \citep{WickramasingheFerrario05}.  The  second
possibility is that  high-field MWDs are produced in  the aftermath of
the merger of two initially  less massive white dwarfs.  This scenario
has been  suggested as a viable mechanism  for producing ultra-massive
white dwarfs ($\MWD>1.1\Msolar$), and  naturally explains why MWDs are
more massive  than their field counterparts.  This  scenario was first
invoked for addressing the hot  and massive population of white dwarfs
in the  ROSAT survey \citep{Marshetal97}, and recently  to account for
the   high  mass   peak   in  the   white   dwarf  mass   distribution
\citep{Kepleretal07}. Since  the incidence of magnetism  is higher for
this      population      of      ultra-massive      white      dwarfs
\citep{WickramasingheFerrario00},  the   merger  scenario  provides  a
natural   explanation   for    the   properties   of   several   MWDs,
\Rej\   \citep{Barstowetal95,    Ferrarioetal97,   Vennesetal03}   and
SDSS\,J150746.80+520958.0      \citep{Dobbieetal11}     being     some
representative examples.

There have  been numerous suggestions  that binary evolution  could be
responsible   for   magnetism    in   white   dwarfs.    Specifically,
\citet{Toutetal08}    and    \citet{Nordhausetal11}   discussed    the
possibility  that  turbulence during  a  common  envelope phase  could
generate  strong magnetic  fields.  These studies  were  aimed at
explaining  two  observational  facts.   The  first  one  is  that  no
high-field MWDs have  been detected in detached binary  systems with M
dwarfs  \citep{Liebertetal05},  while the  second  one  is the  higher
incidence  of  magnetism  in  white dwarfs  in  cataclysmic  variables
(approximately $25$\%) with  respect to that of  single stars (approximately $10$\%)
--- see,   for  instance,   \cite{Warner95}.    All  these   motivated
\citet{Toutetal08} to  put forward the hypothesis that  magnetism is a
direct  consequence of  binarity,  and specifically  of the  evolution
during  the   common  envelope  phase   of  a  close   binary  system.
Specifically, \cite{Toutetal08} argued that isolated MWDs have merged,
whereas  those  in  magnetic  cataclysmic variables  have  not.   They
therefore argued that  the fields must have been  generated before the
merger, during the common  envelope phase.  Following the merger there 
would be a period of normal red giant evolution in which the star 
would spin down until the magnetized core would emerge as a slowly 
rotating MWD.  Only when the merger  occurs just when the
envelope  is ejected would  the white  dwarf emerge  rapidly rotating.
However,  a  few  years  later \citet{PotterTout10}  showed  that  the
magnetic fields generated  in this way are not  durable if only simple
diffusion  of the  field into  the  degenerate white  dwarf matter  is
considered, because of the long associated time scales.

There are  other processes that  might produce magnetic  fields during
binary evolution.  One  of such process is the  effect of shear during
the  intense accretion  phase before  ignition in  an  accreting white
dwarf, and also  during the carbon simmering phase  after the ignition
\citep{Piro08}.     By    using     the    physical    reasoning    of
\citet{PiroBildsten07},  \citet{Piro08}  proposed  that  poloidal  and
toroidal fields are produced due  to a Taylor-Spruit dynamo during the
differentially  rotating phase.   Their calculations  show that  for a
white  dwarf with  $M_{\rm WD}  =  1.37 \Msolar$,  $\Omega_{\rm WD}  =
0.1\Omega_{\rm  K}  =  0.67\,\rm{s}^{-1}$, $\dot{M}  =  10^{-7}\Msolar
\rm{yr}^{-1}$ --- where $\Omega_{\rm K}$ is the Keplerian velocity ---
the   shear  layer   produces   a  magnetic   field  with   components
$B_\phi\textbf{$\approx$} 10^8\,$G and $B_r\approx10^5\,$G.

The merger of  two white dwarfs has become an  active area of research
during the  last years. This is  mostly due to the  belief that double
degenerate  mergers  might  be   successful  progenitors  of  Type  Ia
supernovae.     Although   this    hypothesis   is    relatively   old
\citep{IbenTutukov84,Webbink84}, this scenario has recently received a
vigorized  interest  ---  see,  for instance,  \cite{Bloometal12}  and
\cite{SchaeferPagnotta12}.  Other reasons  of interest in these models
are the  possibilities that white  dwarf mergers could  form magnetars
\citep{Kingetal01, Levanetal06} and the detection of dust discs around
white      dwarfs      \citep{KilicRedfield07,     Garcia-Berroetal07,
vonHippeletal07,  Gaensickeetal07,  Gaensickeetal08, Brinkworthetal09}
and   dust   discs   around  MWDs   \citep{Reidetal01,   Dufouretal06,
Farihietal11,  KawkaVennes11} which  might be  indicative of  a former
viscous disc phase following a merger process.

In  essence, modern hydrodynamical  simulations of  the merger  of two
white  dwarfs  show  that   the  less  massive  secondary  is  totally
disrupted. Depending on  the mass ratio of the  binary components this
process  occurs  on  a dynamical  time  scale.   As  a result  of  the
disruption of the  secondary, part of its material  is accreted on the
primary, whereas  the rest  of it forms  a rapidly  rotating Keplerian
disc.  This confirms  the early suggestions \citep{TutukovYungelson79,
NomotoIben85} that a thick disc  should be produced as an intermediary
step following the coalescence in  which high accretion rates could be
avoided and a  SNIa explosion could take place.   All these hypotheses
have  been  recently confirmed  by  the  in-depth  SPH simulations  of
\cite{Guerreroetal04},  \cite{Yoonetal07}, \cite{Loren-Aguilaretal09},
and \cite{Danetal12}. In particular, all these simulations show that a
thermally  supported  shock heated  region  forms  around the  central
object (the undisrupted primary)  and, additionally, they also predict
the formation of a massive thick disc supported by Keplerian rotation.

The accretion  phase after the  initial violent merger phase  has also
been subsequently investigated.   Recent calculations of the long-term
evolution  of the  merger  \citep{Piersantietal03I, Piersantietal03II,
Yoonetal07, vanKerkwijketal10}  show that  a supernova explosion  is a
likely outcome, although in the case of off-centre carbon ignition the
result could  also be  an accretion induced  collapse (AIC) to  form a
neutron  star  \citep{NomotoIben85}.    However,  there  are  multiple
processes  that  influence  the  long-term  evolution  of  the  merger
product.  One of these is the  interaction of the white dwarf with the
thin  disc.  Recently,  \citet{SaioNomoto04} studied  this interaction
through  the   viscous  boundary  layer  relying  on   the  models  of
\cite{Paczynski91}    and   \citet{PophamNarayan91}.     The   current
understanding  is  that  an  initial  accretion stage  nearly  at  the
Eddington    rate   (about $10^{-5}\Msolar\    \rm{yr}^{-1}$)   is
unavoidable.  More recently, another  model has been proposed in which
the  thick  disc  evolves  towards a  spherical,  thermally  supported
envelope due to the shear  induced by the differential rotation within
a short  timescale of $10^4\,$s  \citep{Shenetal12, Schwabetal12}.  In
both scenarios, the formation of ONe white dwarfs or, depending on the
total mass of  the system, of neutron stars via  AIC, are the expected
results. 

Finally, \citet{GarciaBerroetal12}  proposed that the  magnetic fields
are  generated in  the  corona  above the  merger  product through  an
$\alpha\omega$  dynamo,  since  convection and  differential  rotation
exist  simultaneously.   Using  the  results  of  the  simulations  of
\citet{Loren-Aguilaretal09}  they  showed  that  very  large  magnetic
fields can  be generated  in the hot  corona resulting from  a merger.
Moreover, these  magnetic fields do  not diffuse into the  white dwarf
core or the surrounding disc, and remain confined to the outer layers,
with strengths comparable to those observed in MWDs.  This finding has
been   recently  corroborated   by   \citet{Schwabetal12}  using   the
H{\o}iland criterion.  All in all, this means that magnetic fields can
be naturally produced in the merger of two white dwarfs.  However, one
problem  of  this  scenario  is  the excess  of  the  remnant  angular
momentum,    that    is     incompatible    with    observations    of
MWDs. \citet{GarciaBerroetal12}  suggested magneto-dipole radiation as
a possible  solution for this drawback  of the model.   However, it is
likely that magneto-dipole radiation  would not be efficient enough to
slow down  the remnants  of the coalescence  to the  observed rotation
periods  of  MWDs.   Nevertheless,  if  the merger  product  has  high
magnetic dipole  fields and the disc  from the merger  survives, it is
possible  that the  long-term  evolution of  the  system will  involve
angular momentum exchange between the  disc and the star, depending on
the magnetic field strength of the central object.

In this  paper we analyse the  long-term evolution of the  disc and of
the central object resulting from the coalescence of two white dwarfs,
carrying  over  some  elements  from  models  \citep{Chatterjeeetal00,
Alpar01,Ertanetal09} constructed  for supernova fallback  discs around
young  neutron stars  \citep{Wangetal06}.  Our  paper is  organized as
follows.   In Sect.\,2 we  explain the  details of  our model  for the
debris disc plus MWD system, in Sect.\,3 we use the results of the SPH
simulations  of  \cite{Loren-Aguilaretal09}  and  the  fallback  model
presented in Sect.\,2 to compare our simulations with the observations
of MWDs.   Finally, in  Sect.\,4, we summarize  our main  findings, we
discuss their significance and we draw our conclusions.

\section{The debris disc}

We  consider a  freely  expanding Keplerian  disc  around the  central
object produced  during the merger.   This disc comprises the  mass of
the disrupted low-mass companion that has not been incorporated to the
central object, as little mass  is ejected from the system and carries
the excess angular momentum of the progenitor binary system.

\subsection{Evolution of the disc}
\label{model} 

As  mentioned, for  debris discs  the outer  boundary of  the  disc is
free. That  is, it is  not tidally truncated  like discs  in binary
systems but, instead,  it may expand. Such discs  are similar to discs
formed  by tidal  disruption  of stars  by  super-massive black  holes
\citep{Cannizzoetal90}  or  supernova   fallback  discs  around  young
neutron stars, e.g.\  anomalous X-ray pulsars \citep{Chatterjeeetal00,
Alpar01}.  The  model we  employ is similar  to the model  proposed by
\citet{Ertanetal09} which describes the  evolution of the thin disc in
terms of its initial mass  and angular momentum.  This description
of  the evolution  of  the disc  presumes  that the  cause of  angular
momentum    transport   in   the    disc   is    turbulent   viscosity
\citep{ShakuraSunyaev73}     rather    than     hydromagnetic    winds
\citep{BlandfordPayne82}.

The temporal evolution  of the surface mass density,  $\Sigma$, in the
disc is described by a diffusion equation \citep{Pringle81}:
\begin{equation}
\frac{\partial \Sigma }{\partial t}=\frac{3}{r}\frac{\partial }{\partial r}%
\left[ r^{1/2}\frac{\partial }{\partial r}\left( \nu \Sigma r^{1/2}\right) %
\right]   
\label{diffuse}
\end{equation}
where $\Sigma$ is the surface mass density and $\nu $ is the turbulent
kinematic viscosity.   The rest of  the thin disc  structure equations
\citep{Franketal02} can be solved algebraically \citep{Cannizzoetal90}
to obtain a viscosity in  the form $\nu =Cr^{p}\Sigma ^{q}$ where $C$,
$p$ and  $q$ are  constants determined by  the opacity regime  --- see
also  \cite{Ertanetal09}.  For  this viscosity  Eq.~(\ref{diffuse}) is
non-linear   and    has   the   self-similar    solutions   found   by
\citet{Pringle74}.      In   the    first     type    of      solution
\citep{Cannizzoetal90} the  angular momentum  of the disc  is constant
($\dot{J}_{\mathrm{d}}=0$),  but the  mass of  the disc  evolves  as a
power-law  in time.  Specifically, the  mass  flow rate  at the  inner
boundary can be expressed as:
\begin{equation}
\dot{M}=\dot{M}_{0}\left( 1+\frac{t}{t_{0}}\right) ^{-\alpha }, \qquad 
\dot{M}_{0} = (\alpha-1)\frac{M_0}{t_0}
\label{eq:Mdot}
\end{equation}
where $t_{0}$ is the viscous time-scale at some characteristic radius,
$M_0$ the  initial mass of the  disc and $\dot{M}_{0}$  is the initial
accretion rate. The  power-law index $\alpha$ can be  written in terms
of  $p$  and   $q$,  which  are  determined  by   the  opacity  regime
\citep{Cannizzoetal90}.   For  electron  scattering  the  exponent  is
$\alpha=19/16$   and   for    the   bound-free   opacity   regime   is
$\alpha=5/4$.  Hence, the difference  between the  two regimes  is not
large, and similar  results can be found by  changing the initial mass
of  the  disc slightly.  For  our  numerical  simulations we  use  the
bound-free regime  as this  would be the  dominant opacity  source for
most of the lifetime of the system.

The second type of solutions describes a disc in which the mass of the
disc is  constant ($\dot{M}_{\rm d}=0$) while the  angular momentum of
the   disc    increases   by   the   torque   acted    by   the   star
\citep{Pringle91}. Such a disc  would be relevant if the magnetosphere
is rotating faster than the  Keplerian velocity at the inner radius of
the   disc.    This  occurs   in   the   so-called  propeller   regime
\citep{IllarionovSunyaev75} if the mass is retained in the disc rather
than being ejected from the system.

The most important  parameter governing the spin evolution  of MWDs is
the viscous time-scale $t_0$, which  is determined by the initial mass
and  angular  momentum  of  the  disc  \citep{Ertanetal09}.   For  our
purposes, we assume a metal disc in the bound-free opacity regime with
$\alpha_{\rm SS}=0.1$, where  $\alpha_{\rm SS}$ is the Shakura-Sunyaev
viscosity, for which the viscous time-scale for the white dwarf debris
disc is
\begin{equation}
t_0 \simeq 15\,{\rm s}\,\left(\frac{j_0}{10^{18}\,{\rm cm}^2 \, {\rm s}^{-1}}
\right)^{25/7}\left(\frac{M_0}{10^{-1}\,\Msolar}\right)^{-3/7}
\end{equation}
where $j_0=J_0/M_0$ is the total specific angular momentum.  Note that
the  viscous time-scale  depends strongly  on  $j_0$ and  also on  the
opacity.   However, if the  source of  opacity is  electron scattering
instead of bound-free transitions the viscous time-scale is changed at
most by a factor of two.  For a specific angular momentum one order of
magnitude larger  the viscous  time-scale is of  the order  of 1\,day,
which   is   comparable  to   the   $3-14$\,days  estimate   of
\citet{MochkovitchLivio90}.
        
\subsection{Angular momentum evolution of the white dwarf}
\label{torque}

Angular  momentum  exchange  between  a  disc  and  a  star  has  been
extensively studied  since the discovery  of X-ray pulsars,  and these
findings  are nowadays  applied  to many  astrophysical objects,  like
T~Tauri stars and  cataclysmic variables \citep{Warner95}.  The strong
magnetic field of the white dwarf  disrupts the inner part of the disc
which would  otherwise extend to the  surface of the  star.  The inner
radius of the disc, $R_{\rm in}$, is determined by the balance between
the magnetic and  the material stresses.  This is of  the order of the
Alfv\'en radius
\begin{equation}
R_{\rm A}=\left( \frac{\mu _{\ast }^{2}}
{\sqrt{2GM}\dot{M}}\right) ^{2/7} 
\label{alfven}
\end{equation}
which,  in the  case  of  spherical accretion,  is  determined by  the
balance between magnetic and ram pressures \citep{DavidsonOstriker73}.
Here $G$ is the gravitational constant, $\mu _{\ast }$ is the magnetic
moment of the white dwarf and $\dot{M}$ is the accretion rate reaching
the  the inner  radius but  not necessarily  accreting.  Thus,  we use
$R_{\rm in}=\xi R_{\rm A}$ where $\xi $ is a dimensionless factor used
for converting  the Alfv\'{e}n radius  for spherical accretion  to the
inner  radius  in disc  accretion.   In  most  studies ---  see  e.g.,
\cite{Warner95} --- $\xi \approx 0.5$ is usually adopted. Furthermore,
numerical studies have shown that this value is applicable to the case
of  a  dipolar   field  \citep{Longetal05}.   Nevertheless,  when  the
accretion  rate is  so large  that $R_{\rm  in}$ is  smaller  than the
radius of the star, we assume $R_{\rm in}=R_{\rm WD}$.
 
The  way in  which the  star interacts  with the  disc depends  on the
interplay  between its  inner radius,  the corotation  radius, $R_{\rm
co}=(GM/\Omega _{\rm  WD }^2)^{1/3}$ where  the disc is  rotating with
the same  speed as  the star, and  the light cylinder  radius, $R_{\rm
L}=c/\Omega_{\rm WD}$, $c$  being the speed of light.  To model it, it
is     customary     to      define     the     fastness     parameter
\citep{GhoshLamb79,Lamb88,Lamb89} as
\begin{equation}
\omega _{\ast }=\frac{\Omega _{\rm WD }}{\Omega_{\rm K}\left( R_{\rm in}\right)}=
\left( \frac{R_{\rm in}}{R_{\rm co}}\right) ^{3/2},
\end{equation}
where $\Omega_{\rm  K}(R_{\rm in})$ is the  Keplerian angular velocity
at the inner  radius of the disc and $\Omega_{\rm  WD}$ is the angular
velocity of the star. If the  inner radius of the disc is smaller than
the  corotation radius  ($\omega_{\ast} <  1$)  the system  is in  the
accretion regime.  The inflowing matter  at the inner edge of the disc
will reach  the surface of the  star channelled by  the magnetic field
lines.  If the inner disc  radius is larger than the corotation radius
($\omega_{\ast} > 1$) accretion cannot  occur due to the fast rotation
of   the   star.    This   corresponds   to   the   propeller   regime
\citep{IllarionovSunyaev75,wang85}. In  this phase the  disc continues
its viscous evolution, though matter  from the disc cannot be accreted
on the star due to the centrifugal barrier.

The   torque   applied  by   the   disc   to   the  white   dwarf   is
\citep{GhoshLamb79}:
\begin{equation}
N_{\rm d} = n(\omega_{\ast})N_0.
\end{equation}
where $N_0=\Omega_{\rm  K}\left(R_{\rm in}\right) R_{\rm in}^2\dot{M}$
and $n(\omega_{\ast})$  is called the  dimensionless torque. Following
\citet{Alpar01} and \citet{Eksietal05} we employ $n=1-\omega _{\ast}$.
 Consider a ring of mass $\Delta m$ right outside the magnetospheric 
boundary
$R_{\rm in} + \Delta R$ where the flow is Keplerian. The angular momentum
of the ring is $ \Delta m (R_{\rm in}+ \Delta R)^2 \Omega_{\rm K}(R_{\rm 
in}+ \Delta R)$. Interaction with the magnetosphere will bring the 
matter to the angular velocity of the star at the magnetospheric 
boundary. The change in the angular momentum of the material is then
$\Delta \ell = \Delta m (R_{\rm in}+ \Delta R)^2 \Omega_{\rm K}(R_{\rm 
in}+ \Delta R) - \Delta m R_{\rm in}^2 \Omega_{\ast}$. Assuming the 
transition region is narrow ($\Delta R \ll R_{\rm in}$)
and a continuous mass flow we obtain $N = \dot{M} R_{\rm in}^2 \Omega_{\rm 
K}(R_{\rm in}) (1-\omega_{\ast})$.
Hence this prescription is equivalent to assuming that the magnetosphere and the accreting material are colliding
``particles'' and the interaction between them is inelastic.

If  the inner
radius of the disc goes beyond the light cylinder radius, the disc can
no longer torque  the star via the magnetosphere,  $N_{\rm d} =0$.  In
this so-called ejector stage the  star will spin-down via the magnetic
dipole radiation torque
\begin{equation}
N_{\rm mdr} = -\frac{2\mu_{\ast}^2 \sin^2 \beta\Omega_{\rm WD}^3}{3c^3}
\end{equation} 
where  $\beta$  is the  inclination  angle  between  the magnetic  and
rotation axis.

We consider the effect of change in mass and radius of the star, which
may  be  significant  for  those   stars  with  masses  close  to  the
Chandrasekhar  limit. In  particular, the  rotational velocity  of the
star depends on the moment of  inertia of the star, which changes when
a   substantial   amount   of    mass      (of   the   order   of
about 0.1\,\Msolar)  is accreted  from the  disc.  In  the  case of
accretion  near the  Chandrasekhar mass  limit, the  change  in radius
cannot  be   approximated  by  simple  dependencies.    Hence  in  our
calculations  we  used  the  analytical  mass-radius  relationship  of
\citet{Nauenberg72}
\begin{equation}
\begin{split}
R_{\ast }=0.0126\, {\rm R}_{\sun}&\left( \frac{R}{{\rm R}_{\sun}}\right) 
\left(\frac{M}{{\rm M}_{\sun}}\right) ^{-1/3}\\
\times&\left[ 1-\left( \frac{M}{M_{\mathrm{Ch}}}\right)
^{4/3}\right] ^{1/2}
\end{split} 
\end{equation}
where 
\begin{equation}
M_{\rm Ch}=1.435\, M_{\odot }\left( \frac{2}{\mu }\right) ^{2}
\end{equation}
An  important effect  of the  change  in radius  is that  due to  flux
conservation  the strength  of  the magnetic  field  changes, $\Phi  =
B_{\rm WD}R^2_{\rm WD}$.   This is also considered in  our model.  For
the moment of inertia of white dwarfs we used
\begin{equation}
\begin{split}
I_{\ast }=&3.2\times10^{50}\,{\rm g\,cm}^2\,\left(\frac{M_{\ast}}
{\Msolar}\right)^{0.34158}\,\\
\times&\left[1-\left(\frac{M_{\ast}}{1.437\Msolar}\right)^{1.25}\right]^{1.437}
\end{split}
\end{equation}
which we  obtained by fitting numerical results.  

Finally, we introduce a critical temperature below which turbulence in
the disc can not be sustained. We find the effective temperature using
\begin{equation}
T(r) = \left( \frac{3GM\dot{M}}{8\pi r^3 \sigma} \right)^{1/4} 
\left[1-n\left(\frac{R_{\rm in}}{r}\right)^{1/2} \right]^{1/4}.
\label{eq:disc_temp}
\end{equation}
\citep{Franketal02} where  $n \equiv N/\dot{M}\sqrt{GMR_{\rm  in}}$ is
the dimensionless  torque per  unit mass accretion  rate acted  on the
star.

\section{Results}
\label{sec:results}

\subsection{Discs from SPH simulations}

To compare our model with  observations, we simulated the evolution of
the merger remnants  (namely, the white dwarf and  the disc) using the
results  of the  SPH simulations  of  \citet{Loren-Aguilaretal09}.  We
restricted ourselves to those simulations  for which the total mass of
the system does  not exceed \Mch.  The initial  conditions used in our
simulations ---  that is,  the masses of  the central remnant  and the
disc,  the angular  velocity of  the uniformly  rotating star  and the
angular momentum of the disc --- are shown in Table~\ref{tab:SPH}.

\begin{table}
\caption{Initial   values  of  our   simulations,  as   obtained  from
  \citet{Loren-Aguilaretal09}.}
\begin{tabular}{cccccc}
\hline
\hline
Run & $M_{\rm WD}$ & $M_0$ & $\Omega_{\rm WD}$ & $J_0$ \\
($M_{\sun}$+$M_{\sun}$) & ($M_{\sun}$) & ($M_{\sun}$) & (s${}^{-1}$) & 
(g\,cm\,s${}^{-1}$) \\ 
\hline
0.5+0.3 & 0.62 & 0.18 & $0.134$ & $1.40\times10^{50}$ \\  
0.8+0.4 & 0.92 & 0.28 & $0.244$ & $2.65\times10^{50}$ \\ 
0.8+0.6 & 1.10 & 0.30 & $0.401$ & $2.85\times10^{50}$ \\ 
\hline
\hline
\end{tabular}
\label{tab:SPH}
\end{table}

For a  given simulation, we calculated  the spin evolution  of the MWD
for  different   polar  field  strengths.   To   determine  the  total
integration time  we used the  simple cooling law  of \citet{Mestel65}
and the  otherwise typical observed  value of \Teff$\approx$40\,000~K.
The exact value of of the initial core temperature depends on the mass
of the merging components.  However,  its exact value has virtually no
consequences in  the age estimate,  as white dwarfs cool  very rapidly
during  the early phases  of evolution.   Consequently, we  adopted an
initial temperature  of $10^8\,$K, a  representative value of  the SPH
simulations.   The   cooling  ages  obtained  in  this   way  for  the
0.5+0.3\,\Msolar,                  0.8+0.4\,\Msolar\         and
0.8+0.6\,\Msolar\ mergers --- which result in MWDs of masses 0.8, 1.2,
1.4\,\Msolar\ --- were  13, 50 and 100~Myr respectively.   It could be
argued that for high core temperatures neutrino emission would shorten
significantly the cooling times \citep{Althausetal07}.  However, given
the simplicity  of the  approach adopted here  we consider  that these
cooling times are accurate enough.  Moreover, we found that when these
cooling times are compared with those obtained using full evolutionary
calculations  \citep{Althausetal07,Salarisetal10} the  differences are
minor.

In  accordance  with that  explained  in  Sec.~\ref{torque}, the  spin
evolution  of  the  star  is  determined by  the  fastness  parameter,
$\omega_\ast$.  Initially,  the accretion rate  is very large  and the
magnetospheric radius is small.   In this case, the fastness parameter
is smaller than 1. Thus, the inner  radius of the disc is equal to the
stellar  radius  and the  rapid  accretion  of  matter from  the  disc
increases  the angular  momentum of  the star.   Hence,  initially the
remnants of  mergers are  fast rotators, and  they spin at  almost one
third    of   the    critical   angular    velocity   of    the   star
$\omega_\ast\approx0.33$.   This brief  spin-up  stage continues  till
$\omega_\ast> 1$.   At this  point the magnetic  torque from  the disc
starts  to spin  down the  star.   At this  stage the  star can  reach
critical angular velocities if the mass of the white dwarf is close to
\Mch. This is due to the steep dependence of radius on mass near \Mch.
Since spin-down is only  effective for $\omega_\ast>1$, for very large
accretion rates  it might be  possible that the star  reaches critical
rotation  before the  magnetospheric  radius becomes  larger than  the
stellar   radius.   This  was   the  case   for  the   simulations  of
\citet{Piersantietal03I},                    \citet{Piersantietal03II},
\citet{SaioNomoto04},  and  \citet{Yoonetal07}.   In our  calculations
this situation is avoided for the 0.5+0.3\,\Msolar\ and the  0.8+0.4\,
\Msolar\  mergers.  Hence,  in  the  rest of  the  paper we  only
discuss these  calculations.  Additionally, it could  be possible that
the star  might spin-up to a  point in which the  inner radius exceeds
the  light  cylinder  and,  thus,  accretion ceases.   In  this  case,
magnetodipole   spin-down   (a   less   efficient   mechanism)   takes
over. However, in our calculations we do not find this to be the case.

As  the accretion  rate  decreases, the  magnetospheric radius,  hence
$\omega_{\ast}$, increases rapidly.  This results in a stronger torque
on  the  star which  eventually  forces  it  to reach  equilibrium  at
$\omega_\ast \approx 1$.  All our simulations evolve across this phase
of  torque transfer,  and  for all  three  cases there  is a  limiting
magnetic field strength, $B_{\rm lim}$, which determines whether a MWD
will enter  this strong spin-down  phase, given the same  cooling age.
If the  dipolar field strength, $B_p$,  is close to  $B_{\rm lim}$ the
final  period of  the star  is very  sensitive to  the exact  value of
$B_p$.  In particular, near $B_{\rm lim}$ a 10\% difference results in
change by a factor of approximately 2 in the final spin period.  Finally,
after this strong spin down phase, if the magnetic field strengths and
cooling age are adequate,  the system reaches quasi-equilibrium around
$\omega_\ast\la 1$.   This phase is  called  the tracking phase.
The reason  why a quasi-equilibrium phase  is reached is  that as mass
transfer  continues, the  magnetospheric  radius varies  to match  the
corotation radius of the star.  Hence, the star never reaches complete
equilibrium and continues spinning down.

The  aforementioned  evolutionary  phases  are  common  for  the  
0.5+0.3\,\Msolar\  and  the  0.8+0.4\,\Msolar\ simulations,  and  the
final  period  of  the  object  is determined  by  $B_p$,  because  in
combination  with   the  mass   accretion  rate,  it   determines  the
magnetospheric stopping  radius, the fastness parameter,  and thus the
strength of the torque on the  star for a given accretion rate.  Since
the accretion  rate varies with  time, the magnetic field  strength of
the  star determines  how  fast  the tracking  phase  is reached.   In
particular, if the dipolar field  strength is large the tracking phase
is reached earlier, because the associated torques are larger.

\begin{figure}
   \centering
   \includegraphics[width=\columnwidth]{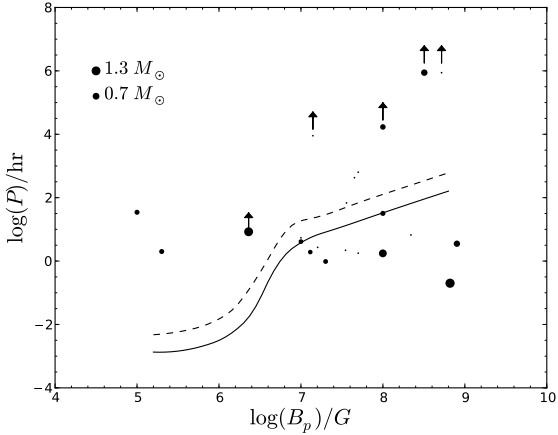}
   \caption{Period versus magnetic field  strength for MWDs with known
     periods, compared to our  results.  The solid line corresponds to
     the case  in which a  1.2\,\Msolar\ MWD is considered,  while the
     dashed line corresponds to the case of a 0.8\,\Msolar\ star.  The
     known  MWDs \citep{Kawkaetal07,Brinkworthetal07}  are represented
     using black circles.  The mass  of the object determines the size
     of the  circle. The  sizes of the  circles for  1.3\,\Msolar\ and
     0.7\,\Msolar\ stars  are shown  in the upper  left corner  of the
     figure.   The stars  with  unknown masses  are represented  using
     small  black dots.  The  data points  with arrows  indicate lower
     limits for the respective periods.}
   \label{fig:PvsB}
\end{figure}

For  the  very early  phases  of  evolution  right after  the  merger,
super-Eddington  accretion  rates are  expected.   This  phase of  the
evolution is  short ($10^4$~yr) with respect to  the evolutionary time
of the observed population of MWDs ($10^7-10^8$~yr), but it determines
the  maximum angular  velocities  attained at  the  beginning of  disc
evolution, due  to the  initial spin-up phase.   However, for  a given
white dwarf and disc angular  momentum, the spin at which the tracking
phase  is  reached  is   determined  by  $B_p$.   Actually,  a  simple
relationship between  $B_p$ and  the final period  of the star  can be
easily computed.   Taking into account that during  the tracking phase
the  angular velocity of  the star  is almost  equal to  the Keplerian
velocity  at  the inner  radius  of the  disc,  and  the accretion  is
modelled  by a  power  law, the  period  of the  star  can be  written
explicitly as:
\begin{eqnarray}
P_{\ast} &=& 3.4\times10^{7}\,{\rm s}\,
\left(\frac{M_{\ast}}{1.3\,\Msolar}\right)^{-10/14}
\left(\frac{B_p}{10^7\,\rm{G}}\right)^{6/7}\nonumber \\ 
&\times&
\left(\frac{R_{\ast}}{4.3\times 10^{-2}\Rsolar}\right)^{18/7}
\left(\frac{J_0}{2.5\times 10^{50}\rm{g\,cm}{}^2\rm{s}{}^{-1}}\right)^{25/28}
\nonumber \\
&\times&
\left(\frac{t}{7.8\times 10^7\,\rm{yr}}\right)^{5/4},
\label{eq:tracking}
\end{eqnarray}
which holds for  $B_p > 10\,$MG. The  results of our calculations
are  shown  in Fig.~\ref{fig:PvsB}.   As  can  be  seen, although  the
periods  obtained in this  way are  more or  less consistent  with the
majority of periods  of the population of MWDs,  which have periods of
days to  weeks --- see  \cite{Kawkaetal07} and references  therein ---
the majority of  the observed spin periods seem to  have a fairly flat
distribution between 1 and  100~hr irrespective of the magnetic field.
On  the contrary,  the models  show  a strong  correlation with  field
strength  in this  range of  periods.  Before discussing this point we 
want to emphasize that not all magnetic white dwarfs have to be the 
result of a merger, and that our model is only aimed at reproducing the 
observed periods of massive white dwarfs. We elaborate on this point 
below.

\subsection{Low-mass discs}
\label{section:tiny}

Up  to this  point  we  have used  an  initial configuration  directly
derived    from   the    results   of    the   SPH    simulations   of
\cite{Loren-Aguilaretal09}. Consequently, we  have assumed that during
the initial phases  of the evolution the central  white dwarf accretes
at  the Eddington  rate, in  accordance  with previous  works ---  see
Sect.~\ref{sec:introduction}.  However,  an alternative model recently
proposed  by \citet{Shenetal12}  assumes that  the disc  disperses its
angular  momentum   due  to  shear  in   the  differentially  rotating
layers. The reasoning of \citet{Shenetal12}  is based on the fact that
radiative cooling is very inefficient. Hence, the material of the disc
cannot  relax to  a thin  disc configuration.   This leads  to entropy
deposition  within the disc  due to  viscous evolution.   The detailed
modelling of this phase was later done by \citet{Schwabetal12} leading
to  viscous time  scales of  $10^4\,$s,  within which  the thick  disc
disperses and  becomes thermally supported  and spherically symmetric.
Actually, \citet{Shenetal12} find:
\begin{equation}
t_{\rm visc} \approx\frac{1}{\alpha_{\rm SS}}
\left(\frac{r_{\rm cyl}}{h}\right)^2
\left(\frac{r_{\rm cyl}^3}{GM_0}\right)^{1/2} \approx
3\times10^4\,\rm{s}
\end{equation}
where $h$ is the thickness of the disc, and $r_{\rm cyl}$ is its size.

\begin{figure}
   \centering
   \includegraphics[width=\columnwidth]{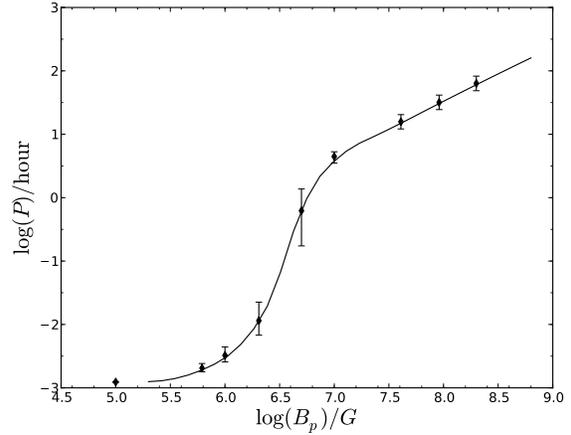}
   \caption{Comparison of the calculations in which the results of SPH
     simulations  are adopted  (solid  line), and  the simulations  in
     which  a   low-mass  disc  was   adopted  (black  dots),   for  a
     1.2\,\Msolar\ MWD.   The error bars represent the  results of the
     simulations with different disc angular momenta.  The upper error
     bars are the  periods obtained when the disc  angular momentum is
     half  of that  resulting from  the SPH  simulations,  whereas the
     lower  periods  result  from  simulations in  which  the  angular
     momentum is  a factor  of two larger  than that of  the SPH
     calculations.}
   \label{fig:compare}
\end{figure}

This model,  however, has  two drawbacks.  The  first one is  that the
model assumes  that no accretion occurs  whatsoever. Consequently, the
evolution  of the  disc  is only  governed  by the  change in  angular
momentum --- see Sect.~\ref{model}. This,  in turn, forces the disc to
evolve changing its angular momentum, as opposed to the case where the
mass is  allowed to be accreted  on the central object,  in which case
the  disc can  expand to  conserve its  angular momentum.   The second
drawback of this model is  the assumption that radiative energy losses
within the disc  are negligible.  A very crude  approximation using an
electron  scattering opacity and  the initial  surface density  of the
disc leads to
\begin{equation}
t_{\rm RL} \approx\frac{h\kappa \Sigma_0}{c} \approx
2\times 10^9\,\rm{s}
\end{equation}
where $\kappa$ is the opacity  of the disc.  This time-scale is orders
of magnitude longer than $t_{\rm visc}$.  However, both $t_{\rm visc}$
and $t_{\rm  RL}$ depend  strongly on $r_{\rm  cyl}$.  As a  matter of
fact  it  turns  out  that  $t_{\rm RL}/t_{\rm  visc}  \propto  r_{\rm
cyl}^{-11/2}$.  Hence, if the disc size increases during the evolution
due to accretion on the central object, radiative cooling could become
efficient and the  disc would become thin. In our  models, a thin disc
reaches  this  size after  $10^2$\,s  if  electron  scattering is  the
dominant source  of opacity. This  is two orders of  magnitude shorter
than $t_{\rm  visc}$.  It is, nevertheless, important  to realize that
up  to  now  we  have   adopted  a  power  law  for  $\dot{M}$.   This
approximation is safe for the later stages of the evolution, when most
of  the   mass  of   the  disc   could  be  lost   if  the   model  of
\citet{Shenetal12} is valid.  Hence,  to estimate the final periods of
the MWDs we  also model these low-mass discs. In  particular, we run a
second set of  simulations in which we adopted the  same total mass of
the system, but we assumed a disc mass $10^{-3}\,\Msolar$ --- the mass
of the disc when the  Eddington accretion phase finishes and accretion
driven by the  evolution of angular momentum takes  over --- while the
rest of  the material was  incorporated into the remnant  white dwarf.
We performed this test for the 0.8+0.4\,\Msolar\ merger --- which
in this case results in a  MWD of about 1.2\,\Msolar.  Our results are
shown in Fig.~\ref{fig:compare}.  As can be seen, the final periods of
these simulations are almost identical to those previously presented.
The reason  for this is that the  final periods reached at  the end of
the  simulation for dipolar  field strengths  below $B_{\rm  lim}$ are
sensitive to the starting angular  momentum of the star, since at this
phase the  star carries most of  the angular momentum  from the merger
process.   However,  for  stars  with dipolar  field  strengths  above
$B_{\rm  lim}$,  the  final  periods  are determined  by  the  angular
momentum of  the disc, see Eq.~(\ref{eq:tracking}).   This equation is
valid for cooling ages longer than about  $10^7\,$yr and $B_p > B_{\rm
lim}\approx 10\,$MG,  which apply for most high-field  MWDs with known
periods.

\begin{figure}
   \includegraphics[width=\columnwidth]{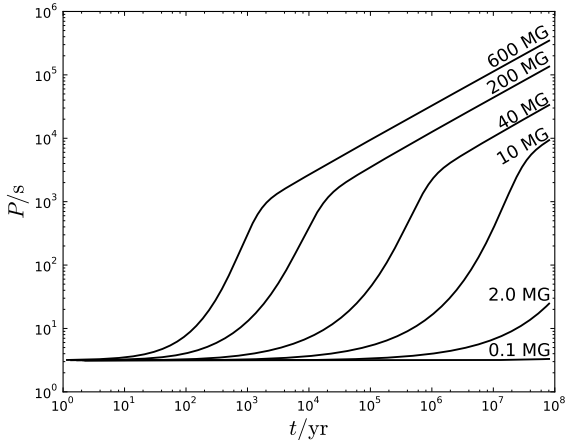}
   \caption{The evolution of the period of a 1.30\,\Msolar\ MWD with a
     disc of  mass $10^{-3}$\,\Msolar\ for  different initial magnetic
     field strengths.}
   \label{fig:Pvst01}
\end{figure}

The angular  momentum of the disc  depends on the mass  of the primary
and the mass difference between the  two components. As can be seen in
Table~\ref{tab:SPH},  the angular  momenta of  the mergers  studied by
\cite{Loren-Aguilaretal09}  differ  by  about  a factor  of  two.   To
simulate arbitrary mass ratios, we made trial runs with different disc
angular momenta  within a  factor of  2 of those  obtained in  the SPH
simulations  ---  see  Fig.~\ref{fig:compare}. Our  calculations  show
that, except  for the case in  which the magnetic  field strengths are
close to $B_{\rm lim}$ --- that is, when the evolution is stopped just
before  the quasi-equilibrium  phase is  reached ---  the  results are
nearly identical to those in which the angular momentum resulting
from the SPH simulations is adopted. In fact, although for the strong
torque transfer phase the periods can differ by an order of magnitude,
for the tracking phase the  maximum difference in the computed periods
is  approximately 20\%.

\subsection{Some insight from \Rej\ and \euve}

To compare our results with  the properties of two known ultra-massive
MWDs  which are  supposed to  be  the result  of a  merger ---  namely
\Rej\  and \euve\, ---  we followed  the evolution  of a  remnant with
their approximate  mass, $1.30\,\Msolar$. For this simulation we 
adopted the following initial conditions. The initial angular momentum
was $J_0=2.5\times 10^{50}$~g~cm~s$^{-1}$, whereas the initial angular
velocity was $\Omega_{\rm K}/3$.  We remind the reader that, according
to  that  discussed  in  Sect.~\ref{section:tiny}, the  value  of  the
initial  angular momentum  plays a  limited role,  whilst  the initial
value of  the angular  velocity does  not have a  large impact  in our
results as long as we only consider the values of the periods obtained
during the  tracking phase.   The observed period  of \Rej\  is 721~s
\citep{Barstowetal95,Ferrarioetal97}, whilst in  the case of \euve\ no
conclusive   evidence   of   rotation    has   been   found   so   far
\citep{Schmidtetal92}.  However, for this last object the instrumental
timing limit  of 8.4~h can be adopted  as a lower limit  to the period
\citep{Shtoletal97}. The  magnetic field  strength of  \Rej\ is
between  170~MG   and  660~MG  \citep{Ferrarioetal97,  Burleighetal99,
Vennesetal03,   FerrarioWickramasinghe05rot},   the   magnetic   field
strength     of     \euve\    is     estimated     to    be     2.3~MG
\citep{Schmidtetal92}. For  these specific simulations  we stopped the
simulations   at   the   cooling   age   for   which   $T_{\rm   eff}
\approx40\,000\,$K,   a  value   consistent  with   the  observational
determination   for    \Rej\   \citep{Barstowetal95,   Ferrarioetal97,
Vennesetal03}  and  30\,000~K,  the  effective temperature  of  \euve.
However, we  stress that  if $T_{\rm eff}=40\,000\,K$  is taken  as an
upper limit, the final ages for both objects are lower limits ($t_{\rm
cool}  \approx 10^8\,\rm{yr}$) and  consequently the  computed periods
are also lower  limits.  The evolution of the period  as a function of
time     is      shown     in     Fig.~\ref{fig:Pvst01}.      Clearly,
Fig.~\ref{fig:Pvst01}  reveals that  our simulations  cannot reproduce
the  observed properties of  these objects,  and that  another braking
mechanism must operate.  In particular, for \Rej\ our model predicts a
period  which is  much  longer than  that observationally  determined,
while for \euve, which has a smaller magnetic field, we obtain periods
of  the  order  of   $10^2\,$s,  much  shorter  than  expected.   More
generally,  for dipole  magnetic field  strengths larger  than $B_{\rm
lim} \approx 10\,$MG --- namely, for the case of \Rej\ --- the periods
are  longer than  about $10^4\,$s,  while  for field  strengths
smaller  than this  value ---  which  is the  case of  \euve\ ---  the
spin-down mechanisms implemented in our model are not adequate either.
Hence,  our  simulations show  that  neither  viscous dissipation  nor
magneto-dipole spin-down  are able to explain the  observed periods of
these  objects, confirming  our previous  finding  that  other
alternatives must be sought.

\begin{figure}
\includegraphics[width=\columnwidth]{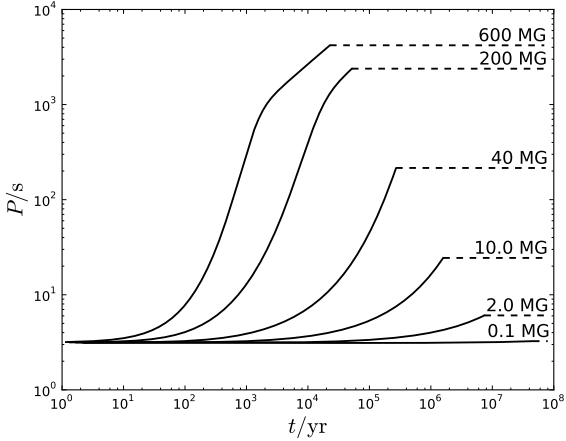}
  \caption{The evolution  of the period for a  1.30\,\Msolar\ MWD. The
    simulations  are  identical  to  those  of  Fig.~\ref{fig:Pvst01},
    except  for  the  inclusion  of  turbulence  turn-off  at  $T_{\rm
    p}=1\,000\,$K.  The  dashed lines  correspond to the  phase during
    which the  torque transfer from  the disc becomes  ineffective and
    magneto-dipole spin down takes over.}
  \label{fig:Pvst02}
\end{figure}

\subsection{Models with turbulence turn-off}

An  important question  to be  analysed is  whether turbulence  in the
accretion disc can be sustained efficiently during the entire lifetime
of  MWDs.  This  question is  relevant when  the lifetimes  of neutron
stars with fallback  discs and MWDs are compared.   In the former case
the  evolutionary  ages  are  typically  $10^4-10^5$\,yr,  whilst  the
lifetimes of MWDs are three  to four orders of magnitude longer. Thus,
this question is an important factor to be considered.

Whether  the disc  is turbulent  or not  is mainly  determined  by the
degree of ionisation of the  disc material.  That is, the existence of
turbulence depends  on the temperature, which, in  turn, is determined
by the accretion rate --- see Eq.~(\ref{eq:disc_temp}). The ionisation
stability  of  accretion discs  has  been  investigated thoroughly  by
\cite{Menouetal01}.    In  their  work   they  computed   the  thermal
equilibrium  (the so-called  S-curves)  of thin  accretion discs  with
different chemical  compositions, and showed that for  thin discs made
of C,  thermal equilibrium is  reached near $5\,000\,$K, and  that for
all   compositions   recombination   always  occurs   below approximately
$1\,000\,$K. For  the conditions found  during the final stages  of our
simulations  these temperatures  are relatively  high. However,  it is
known  that  these  discs  can  be partially  ionised  at  even  lower
temperatures, due to several other processes, like cosmic ray heating,
collisions  and   charged-particle  mobility.   \citet{InutsukaSano05}
carefully studied all these effects and reached the conclusion that at
temperatures below about $1\,000\,$K, magneto-rotational instability
(MRI)   could  be   still  effective.    Building  upon   their  work,
\citet{Ertanetal09}  modelled  the  evolution  of  debris  discs  
assuming  that the  differentially-rotating  layers with  temperatures
smaller than  a critical temperature,  $T_{\rm p}$, do not  change the
angular momentum of the star.  Moreover, they showed that when $T_{\rm
p}\approx  100\,$K  is   adopted  the  observations  are  consistently
reproduced.

Contrary to the modelling  of \citet{Ertanetal09}, in which turbulence
is switched-off layer by layer, we applied a simpler procedure to test
for  how long  the disk  would be  turbulent. For  each time  step, we
calculated the temperature at  $r=10 R_{\rm in}$.  If this temperature
was smaller than $T_{\rm p} = 1\,000\,$K, we assumed that the disc was
not  turbulent   and  accretion  was  switched-off.    Thus,  the  MWD
immediately enters in the so-called  ejector phase and spins down only
due to magneto-dipole radiation.  To compare these calculations to the
case of  \Rej\ and \euve\  we applied this  procedure to the  model of
$1.3\,\Msolar$  described in  Sec.~\ref{section:tiny}.  We  found that
although  the  MRI is  long-lived  ( about $10^5-10^7\,$yr), the  disc
becomes   inactive   at   the    current   ages   of   these   objects
(about $10^8\,$yr). Moreover, since  the magneto-dipole torque is not
as strong  as the magnetic torque  from the disc, the  final period is
determined by  the time at which  MRI is turned off,  which depends on
the     accretion    rate,    and     can    be     determined    from
Eq.~(\ref{eq:disc_temp}):

\begin{eqnarray}
P_{\ast} &=& 1.3\times10^{3}\,{\rm s}\nonumber 
\left(\frac{M_{\ast}}{1.3\,\Msolar}\right)^{-10/28}\\
&\times&\left(\frac{\dot{M}_{\ast}}{2.2\times10^{-11}\,\Msolar\,\rm{yr}^{-1}}
\right)^{1/28} \nonumber \\
&\times&\left(\frac{B_p}{10^8\,\rm{MG}}\right)^{3/7}
\left(\frac{R_{\ast}}{4.3\times10^{-3}\Rsolar}\right)^{-9/7}\nonumber \\
&\times&\left(\frac{T_{\rm p}}{1\,000\,{\rm K}{}}\right).
\end{eqnarray}

As can be seen   in Fig.\,\ref{fig:Pvst02}, the accretion rate at
the  time of  MRI deactivation  depends also  in the  strength  of the
magnetic field.  If  the torque exerted by the  disc is strong enough,
the  disc becomes  inactive after  reaching the  tracking  phase.  For
$T_{\rm p}  = 1\,000\,$K, for white  dwarfs of masses  between 0.8 and
1.3~\Msolar\ this  occurs for magnetic fields around  90\,MG.  In this
case, the  final period  of the  star is determined  by $B_p$  and the
critical temperature of  MRI turn-off.  Using Eq.~(\ref{eq:disc_temp})
once more, the value of $\dot{M}$ at the point at which $T_{\rm p}$ is
reached can be determined, and employing this value, the period at the
time of MRI turn-off can be computed:

\begin{eqnarray}
P_{\ast} &=& 1.3\times10^{3}\,{\rm s}\,
\left(\frac{M_{\ast}}{1.3\,\Msolar}\right)^{-5/13} \nonumber\\
&\times&\left(\frac{R_{\ast}}{4.3\times10^{-3}\Rsolar}\right)^{18/13} \nonumber\\
&\times&\left(\frac{B_p}{10^8\,\rm{G}}\right)^{6/13} 
\left(\frac{T_{\rm p}}{1\,000\,{\rm K}{}}\right)^{-12/13}.
\end{eqnarray}

This  expression shows  that  for MWDs  with  field strengths  between
50\,MG  and 1\,GG,  and $T_{\rm  p}$ ranging  from   $300\,$K to
1\,000\,K,  the final  periods reached  are around  $10^3-10^4$~s (see
also Fig.\,\ref{fig:Pvst02}).

\begin{figure}
   \centering
   \includegraphics[width=\columnwidth]{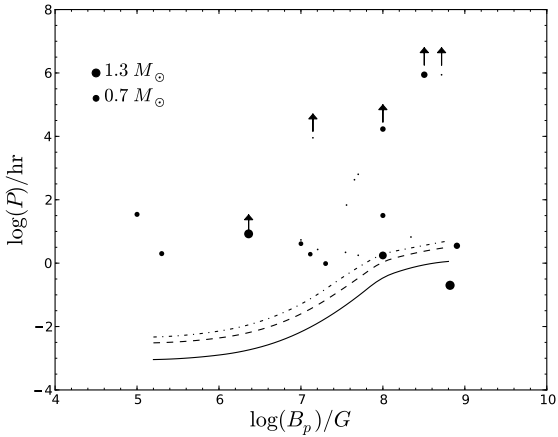}
   \caption{Same as  Fig.~\ref{fig:PvsB} for the  simulations in which
     turbulence is  switched-off at $T_{\rm  p}=1\,000\,$K.  The solid
     line represents the spin evolution  of a MWD of 1.3\,\Msolar, the
     dashed line  that of a  MWD of mass 1.0\,\Msolar,  and dot-dashed
     one  corresponds  to  the  case  in a  mass  of  0.8\,\Msolar  is
     adopted.}
   \label{fig:PvsB2}
\end{figure}

\begin{table*}
\caption{Comparison of  the simulated  and observational data,  when a
  an MRI turn-off temperature $T_{\rm p} = 1\,000\,$K is adopted.}
\begin{tabular}{cccccccc}
\hline
\hline
WD & Other names & $M_{\rm WD}$ & $B$ & $B_p$ & $P_{\rm obs}$ & $P_{\rm sim}$ & Refs.\\
   &             & ($M_{\sun}$) & (MG) & (MG) & (min) & (min) & \\ 
\hline
0325$-$857 & \Rej & 1.35 & 170$-$660  & 200 & 12.08 & 28 & 1,2,3\\  
1015+014   & \PGa & 1.15 & 50$-$90    &  90 & $98.84^{+0.14}_{-0.07}$ & 36 & 4,5,6\\ 

1031+234   & \PGb & 0.93 & 200$-$1000 & 500 &$211.8\pm3.0$ & 205 & 4,6,7 \\ 
\hline
\hline
\end{tabular}
\\(1) \citet{Barstowetal95}, 
  (2) \citet{Ferrarioetal97}, 
  (3) \citet{Kulebietal10}, 
  (4) \citet{Liebertetal03}, 
\\(5) \citet{Euchneretal06}, 
  (6) \citet{Brinkworthetal07}, 
  (7) \citet{Schmidtetal86} 
\label{tab:comparison}
\end{table*}

 In   Fig.~\ref{fig:PvsB2}  we  compare  the   results  of  these
calculations with the observational data. This is done for three white
dwarfs masses,  0.8\,\Msolar, 1.0\,\Msolar\ and  1.3\,\Msolar.  As can
be  seen in  this figure,  now the  computed periods  are considerably
shorter,  and match  better the  periods  of MWDs  with known  masses.
However,  in  order  to  perform  a  meaningful  comparison  with  the
observations only  those MWDs which are  likely to be the  result of a
merger need to  be considered.  Thus, we removed  all MWDs with masses
smaller  than 0.8\,\Msolar.   Of these  MWDs six  have  known periods.
Three of them  are well known slow rotators,  namely \Grw, G\,240$-$72
and  \euve, for which  only lower  limits to  their periods  have been
determined.  The remaining  three are \PGa, \PGb\ and  \Rej, which all
have magnetic  fields above $50\,$MG, and  periods of the  order of an
hour or less.  We simulated the secular evolution of the spin of these
three  MWDs,  using the  observationally  determined  values of  their
masses      and     magnetic      field     strengths      ---     see
Table\,\ref{tab:comparison}.   In addition  to the  properties  of the
disc,  the most  important input  in  our simulations  is the  dipolar
magnetic  field  strength.   Specifically,  it  has  been  shown  that
probably  \Rej\  \citep{Barstowetal95, Ferrarioetal97,  Vennesetal03},
\PGb\  \citep{Schmidtetal86}   and  \PGa\  \citep{Euchneretal06}  have
dipolar  field  components   considerably  smaller  than  the  maximum
observed field strength.  Thus, for them we adopt the values listed in
the fifth  column of  Table~\ref{tab:comparison}, which in  general are
somewhat  smaller  than the  observed  surface  field strengths.   The
results of  our simulations are  shown in Table\,\ref{tab:comparison}.
We note that the simulated periods for these three MWDs (listed in the
seventh  column of this  table) compare  favourably with  the observed
periods (sixth column).  In  particular, for \Rej\ the computed period
is  a factor  of two  longer, while  for \PGa\  is a  factor  of three
shorter, and for  \PGb\ we find a nice  agreement.  All this indicates
that  these objects  were  most  likely originated  in  a merger,  and
subsequently  lost their  angular momentum  due to  the magnetospheric
interaction with a remnant disc.

\subsection{Winds}
\label{section:wind}

It is currently understood that  the initial rapid accretion phase can
not be  approximated by cold accretion,  and that a  hot region around
the    central    MWD    would    be    formed    ---    see,    e.g.,
\cite{vanKerkwijketal10}.   In order  to qualitatively  understand the
implications of this  for the angular momentum evolution  of a MWD, we
built a  model in which  the spin  down is due  to the outflow  of the
material rather than to accretion onto the star. In this prescription,
mass transfer between  the disc and the star  occurs, but the material
is not bound to the system, and it is ejected.

The  induced torques  on the  star can  be then  approximated  by wind
torques \citep{WeberDavis67}:
\begin{equation}
N_{\rm w} = -\eta\dot{M}_{\rm w}\Omega_\ast R_{\rm A}^2
\end{equation}
where $\dot{M}_{\rm  w}$ is the wind  mass loss rate, and  $\eta$ is a
parameter which depends on the geometry  of the mass loss, which for a
spherically symmetric  wind is $2/3$. In this  expression the Alfv\'en
radius $R_{\rm A}$  is computed as the distance  at which the Alfv\'en
speed equals the  wind velocity, rather than the  incoming velocity of
the accreting  material, as  in Eq.~(\ref{alfven}). Note  that $R_{\rm
A}$ depends on the geometry of the magnetic field, and simulations are
necessary    for    precise   values    ---    see,   for    instance,
\cite{MattPudritz05,MattPudritz08}.    However,   for   the  sake   of
simplicity, we assume  that the mass-loss rate  follows the same power
law  given  by  Eq.~(\ref{eq:Mdot})  and  that  $R_{\rm  A}$  is  that
corresponding to disc accretion.

\begin{figure}
\includegraphics[width=\columnwidth]{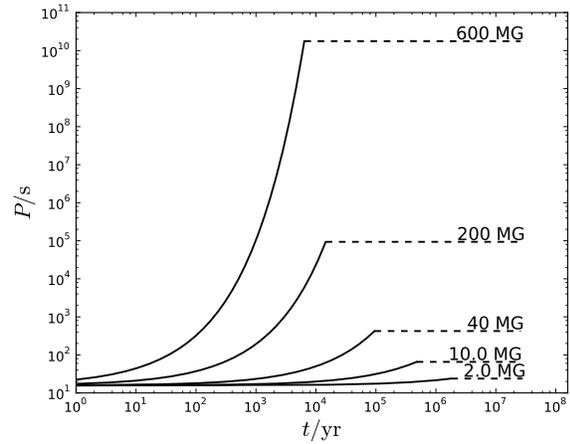}
   \caption{The evolution of the period  of a 1.1\,\Msolar\ MWD due to
     torques  from  the  wind  with  turbulence  turn-off  at  $T_{\rm
       p}=1\,000\,$K.  The initial conditions  were those  obtained in
     the SPH  simulation of the  merger of two  0.8+0.6\,\Msolar white
     dwarfs (see Table~\ref{tab:SPH}).}
   \label{fig:Pvst03}
\end{figure}

We find that  during the early phases of the  evolution the spin rates
are not  strongly affected, since  the mass-loss rates are  very high,
and $R_A$ is  small. However, as the mass-loss  rate decreases $R_{\rm
A}$ becomes larger, and a strong  torque on the object is exerted.  In
fact, the  most apparent qualitative  difference between the  wind and
the disc spin down scenarios is that in the first one the torque spins
down  the MWD  most  effectively  at relatively  early  phases of  its
evolution, before reaching a  quasi-equilibrium, whereas in the second
one the wind spin down is ineffective during the initial phases of the
evolution, and as the mass-loss rate drops, the torque rapidly becomes
stronger.   Moreover,  in  this  case  the  period  evolution  can  be
calculated analytically, once  it is taken into account  that the only
time-dependent quantity now is $\dot M$:
\begin{equation}
\ln\left(\frac{P}{P_{0}}\right) \propto \frac{\kappa}{I_\ast} 
M_\ast^{2/7}R_\ast^{24/7}M_0^{3/7}t_0^{3/28}B^{8/7}t^{13/28} 
\end{equation}
where  $P_{0}$ is  the initial  period of  the star.   This expression
indicates that  the final  periods of MWDs  are very sensitive  to the
values of the stellar parameters,  and that depending on their precise
value large spin downs might be obtained.

We  are interested  in estimating  the  time needed  to reach  periods
similar to that of  \Rej, i.e.  $P_{\ast}\approx10^{4}\,$s.  Assuming that the
initial period is about $10$\,s, the corresponding time-scale is:

\begin{eqnarray}
t_{\rm MWD} &\approx& 2.35\times10^5{\rm yr}\, 
\left(\frac{2/3}{\eta}\frac{I_\ast}
{5.4\times10^{49}\,{\rm g\,cm}^2}\right)^{28/13} \nonumber\\
&\times & \left(\frac{M_\ast}{1.1\,\Msolar}\right)^{-8/13} 
 \left(\frac{R_\ast}{7\times10^{-3}\,\Rsolar}\right)^{-56/13}\nonumber \\
&\times & \left(\frac{M_0}{0.3\,\Msolar}\right)^{-12/13} 
 \left(\frac{t_0}{1\,\rm{s}}\right)^{-3/13}
\left(\frac{B}{10^8\,\rm{G}}\right)^{-32/13}
\end{eqnarray}

For  strong   dipolar  field  strengths   periods  of  the   order  of
$10^3\,$seconds  are  reached in  time-scales  much  smaller than  the
cooling  age   of  \Rej\  and  similar   objects.   Additionally,  our
simulations  with turbulence turn-off  show that  for MWDs  with field
strengths larger than $200$\,MG the wind spin down is effective long enough to
induce strong  spin down ---  see Fig.~\ref{fig:Pvst03}. We  also note
that in this  case the resulting periods are similar  to those of \Grw
and GD~229, which $>100$~yr --- see \citet{Kawkaetal07} and references
therein.   On   the  contrary,   for  field  strengths   smaller  than
about $50\,$MG,  the wind torques  are ineffective,  yielding periods
smaller than  a few minutes ($P_{\ast}<3\times10^2\,$s)  which are unrealistic
for the observed periods of the MWDs.

\section{Conclusions}

We  have studied  the  long-term evolution  of  magnetic white  dwarfs
surrounded by debris discs, which are thought to be the final products
of the  merger of two  degenerate cores.  To  do so we built  a simple
model  to follow  the  coupling  between the  central  object and  the
surrounding  disc.  This  model  is  similar to  those  used to  study
supernova fallback discs around young neutron stars,  and presumes
that  the evolution  of  the  disc is  dominated  by angular  momentum
transport by turbulent viscosity.  Contrary to previous studies of the
evolution of discs  around white dwarfs, our model  considers a freely
expanding Keplerian  disc.  Using this  model we found that  for these
systems there is a relation between the mass, the rotation period, and
the strength of the magnetic field.

In a first  set of simulations we studied the  evolution of the merger
remnants using  the results of  detailed SPH simulations  and assuming
that the  entire disc was turbulent.   For these models  we found that
the final  spin of the white dwarf  is given by a  simple relation ---
see Eq.~(\ref{eq:tracking}). However, the periods obtained in this way
are somewhat larger than  the observed ones.  Consequently, we studied
other alternatives. In  particular, in a second set  of simulations we
computed a set  models in which we assumed ---  in accordance with the
studies of \cite{Yoonetal07, vanKerkwijketal10, Shenetal12} --- that the
cold accretion assumption  would not be applicable to  the post merger
systems.  Following  the   suggestion  of  \citet{Nordhausetal11},  we
simulated the case  in which the mass transferred  from the dics would
end up  as an outflow, and the  spin down is governed  by the magnetic
interaction with it. We found that, nevertheless, the periods obtained
in this second set of models are  again too large. We also varied by a
factor of  two the initial angular  momentum of the disc,  and we also
found  that  the resulting  periods  do  not  differ much  from  those
obtained using the results of the  SPH simulations. All this led us to
analyse  whether turbulence  in the  accretion disc  can  be sustained
efficiently during the entire lifetime of MWDs. This depends mainly on
the degree of ionisation of  the disc material.  According to the most
recent studies we assumed  that turbulence ceases when the temperature
of the disc is smaller than  a certain threshold, for which we adopted
$T_{\rm  p}=1\,000$~K  \citep{InutsukaSano05}.   When  this  model  is
adopted our  calculations reproduce  the observed properties  of three
MWDs with  known masses and  periods in the solar  neighbourhood which
are massive enough --- namely, with masses larger than $0.8\, \Msolar$
--- to be considered  the result of white dwarf  binary mergers (\Rej,
\PGa,  and \PGb).  These  findings support  the hypothesis  that these
white dwarfs  are the remnants of double-degenerate  mergers that have
gone  through magnetopsheric  interaction with  a remnant  disc.  More
specifically, it has been known for a long time that \Rej, the fastest
rotator within the MWD population, spins slower than what it should be
expected for  a double white  dwarf merger.  Our model  reproduces the
rotation period of this  object, which was previously unaccounted for.
However, we also find that  our model cannot explain the properties of
the rest of  the population of massive MWDs,  which is probably formed
by multiple evolutionary channels \citet{FerrarioWickramasinghe05rot}.
In  particular, there are  three massive  MWDs (\Grw,  G\,240$-$72 and
\euve)  which  are  slow   rotators,  and  probably  have  experienced
spin-down during the red giant phase  either as a single star of after
a  core merger  during  the  common envelope  phase,  as suggested  by
\cite{Toutetal08}.

\cite{FerrarioWickramasinghe05rot}     conducted     very     thorough
observational investigations  of the  properties of the  population of
MWDs and  resolved that within  it there are three  different magnetic
field  and  period  intervals,   which  point  out  towards  different
evolutionary  origins.  Specifically, they  proposed the  existence of
three  distinct groups  of stars.   The first  group corresponds  to a
population  of  magnetized  slow  rotators ---  with  typical  periods
ranging from $P\approx50\,$yr to  $100$\,yr --- which most likely have
evolved from isolated progenitors.  To  the best of our knowledge only
one  qualitative  study  of   the  evolution  of  rotating  MWDs  
originating   from  single   stars   has  been   performed  so   far.
Specifically, \citet{Suijsetal08} estimated the effect of magnetism on
the core rotation  and concluded that the presence  of magnetic fields
leads to slower  rotating cores.  However, their results  do not fully
account  for the  long rotation  periods  of some  MWDs, namely  those
belonging to  the tail of  the distribution (with periods  longer than
about $100\,$yr).   For these stars they proposed  that spin-down by
stellar  winds might  be an  effective alternative.   However, a
binary  origin for this  group of  stars cannot  be discarded,  as the
results  of \cite{Toutetal08}  show that  a merger  during  the common
envelope phase can result as well  in a slowly rotating MWD.  A second
group  o stars  is  that  of strongly  magnetized  fast rotators  ($P
\approx 700$\,s),  which probably  carry the remnant  angular momentum
from  the double-degenerate  merger.   Finally, there  exists a  third
group of rotators  with periods ranging from hours  to days, which may
have a  mixed origin.  Additionally,  they found that the  majority of
MWDs have  periods between a few hours  to a few weeks,  and also that
the known slow rotators are  not all necessarily highly magnetic (e.g.
\euve).    However,   although    the   overall   picture   drawn   by
\citet{FerrarioWickramasinghe05rot}  may not be  significantly altered
by our calculations, our results indicate that the distinction between
all three  groups may  not be so  clear ---  see Fig.~\ref{fig:PvsB2}.
Moreover,  our  results  show  that  slow rotation  does  not  provide
sufficient  evidence  to  discard  a  binary  evolutionary  origin  of
individual MWDs,  since also the  dipole field strength, the  mass and
the age of the specific star need to be considered.

Additionally, our  scenario predicts the existence of  a population of
MWDs surrounded  by dust discs.  Observational searches  of dust discs
around two  massive MWDs (\Rej\  and \euve) have been  performed using
the {\sl Spitzer}  telescope, but no discs  have been detected so
far \citep{Hansenetal06}.   However, the lack of  detection of debris
discs around  high-field MWDs (those  with $B>2\,$MG) might be  due to
the  diffuse  structure  resulting  from  their  large  magnetospheric
stopping radii --- a consequence of the large magnetic pressure of the
remnant  of  the merger  ---  which,  in  turn, produces  an  extended
disc. More generally, our models imply that the  detectability of
a remnant disc  depends not only on  the mass of the MWD,  but also on
its  magnetic  field strength  and  on  its  period.  Clearly,  deeper
observations  are needed  to confirm  the existence  of  these diffuse
discs  around MWDs.  In  this regard  it is  important to  mention the
recent discovery of an unveiled  population of MWDs surrounded by dust
discs.      NLTT~10480     \citep{KawkaVennes11}    and     G77$-$5018
\citep{Farihietal11}  are examples of  MWDs with  significant infrared
excesses, which  could be attributed  to the existence of  debris disc
around them.

Finally, we would like to  mention that one important hindrance of our
modelling is the difference between the geometry of the magnetic field
assumed in  our simulations and  that of the observed  MWD population.
Multi-phase    spectro-polarimetric   analysis   \citep{Euchneretal02,
Euchneretal05} and single  phase spectral studies \citep{Kulebietal09}
of a  large number of  MWDs have showed  that their field  geometry is
more  complicated  than a  simple  dipole, as  it  is  assumed in  our
simplified model.  This is an important issue as numerical simulations
show  that  the  geometry  of  the magnetic  field  is  important  for
determining the  magnetospheric stopping  radius of an  accreting star
\citep{Longetal07,   Longetal08,  Longetal12}.    For   the  case   of
\Rej\   and \PGb\ --- two of the specific  white dwarfs analysed
here --- we  addressed this issue by assuming  dipolar field strengths
smaller than the total measured  value of the magnetic field, a simple
but  effective  procedure.  Thus,  future  studies  in this  direction
should adopt more realistic magnetic configurations.

\section*{Acknowledgments}
This work  was partially supported  by MCINN grant  AYA2011--23102, by
the European  Union FEDER  funds, and by  the ESF  EUROGENESIS project
(grant EUI2009-04167). We  would  like  to  thank  Aldo  Serenelli for 
many fruitful discussions during the course of this research. 

\bibliographystyle{mn2e}
\bibliography{./fall.bbl}

\end{document}